\documentstyle[prl,aps,psfig,floats]{revtex}

\begin{document}

\tighten
\firstfigfalse
\twocolumn[\hsize\textwidth\columnwidth\hsize\csname@twocolumnfalse\endcsname

\title{Signatures of the Tricritical Point in QCD}
\author{M. Stephanov$^1$, K. Rajagopal$^2$ and  E. Shuryak$^3$}
\address{
$^1$ Institute for Theoretical Physics, State University of New York,
Stony Brook, NY 11794-3840\\
$^2$ Center for Theoretical Physics, Massachusetts Institute of Technology,
Cambridge, MA 02139\\
$^3$ Department of Physics and Astronomy, State University of New York, 
     Stony Brook, NY 11794-3800
}

%\date{DRAFT \today\ 
%{\tiny\verb$Id: sig.tex,v 2.4 1999/03/08 23:29:03 misha Exp $}}
\date{June 1, 1998; ITP-SB-98-39, MIT-CTP-2748, SUNY-NTG-98-17}
\maketitle

\begin{abstract} 
Several approaches to QCD with two {\em massless} quarks at finite
temperature $T$ and baryon chemical potential $\mu$ suggest the
existence of a tricritical point on the boundary of the phase with
spontaneously broken chiral symmetry.  In QCD with {\em massive}
quarks there is then a critical point at the end of a first order
transition line.  We discuss possible experimental signatures of this
point, which provide information about its location and properties.  We
propose a combination of event-by-event observables, including
suppressed fluctuations in $T$ and $\mu$ and, simultaneously, enhanced
fluctuations in the multiplicity of soft pions.
\end{abstract}  
\bigskip
]
%\pacs{}
%\narrowtext   

In QCD with two massless
quarks, a spontaneously broken chiral symmetry is restored at 
finite temperature. It can be argued \cite{piswil,rajreview}
that this phase transition is likely second order 
and belongs to the universality class of $O(4)$ spin models in 3 dimensions.
If this transition is indeed
second order, QCD with two quarks of nonzero mass has
only a smooth crossover as a function of $T$. 
Although not yet firmly established, this picture is consistent
with present lattice simulations  and many models. 

At zero $T$ several models 
suggest \cite{njl,steph,colorsuperconductor,bergesraj,stephetal} that
the chiral symmetry restoration transition at finite $\mu$ is {\it first}
order.  Assuming that this is the case in QCD, one can easily argue
that there is a tricritical point in 
the $T\mu$ phase diagram, where the transition changes from 
first to second order. 
The nature of this point can be understood
by considering the Landau-Ginzburg effective potential for the
order parameter of chiral symmetry breaking, 
$\phi=(\sigma,\bbox{\pi})\sim\langle\bar\psi\psi\rangle$:
\begin{equation}
\Omega(\phi) = a\phi^2 + b(\phi^2)^2 + c(\phi^2)^3.
\label{phi6}
\end{equation}
The coefficients $a$, $b$, and $c>0$ are functions of $\mu$ and $T$.
The second order phase transition line described by $a=0$ at $b>0$
becomes first order when $b$ changes sign. The critical properties
of this point can be inferred from universality
\cite{bergesraj,stephetal}, and the exponents are as in the 
mean field theory (\ref{phi6}).

\begin{figure}[htb]
\centerline{\psfig{file=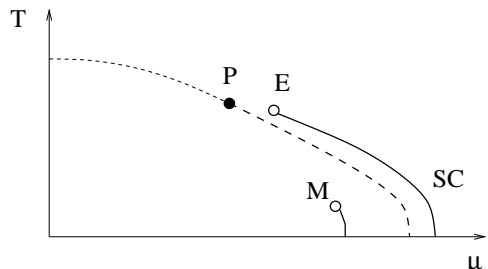,width=2.5in}}
\smallskip
\caption[]{The schematic phase diagram of QCD. The dashed lines
represent the boundary of the phase with spontaneously broken
chiral symmetry in QCD with 2 massless quarks. The point P
is tricritical. The solid line with critical end-point E
is the line of first order transitions in QCD with 2 quarks
of small mass. The point M is the end-point of the
nuclear liquid-gas transition probed in multifragmentation
experiments. The superconducting phase of QCD 
\cite{bailin,colorsuperconductor,bergesraj}, 
marked SC, is not relevant to our discussion.
} 
\label{fig:pd}
\end{figure}

In real QCD with nonzero quark masses 
the second order phase transition becomes a crossover and
the tricritical point becomes
a critical (second order) end-point of a 
first order phase transition line.
Universality arguments \cite{lawrie,stephetal}
also predict that the end-point E in QCD with small quark masses
is shifted with respect to the tricritical point P towards larger $\mu$ as 
shown in Fig. \ref{fig:pd}.
It can also be argued \cite{bergesraj,stephetal}
that the point E is in the universality class of the 
Ising model in 3 dimensions, because the $\sigma$ is the
only field which becomes massless at this point. (The pions
remain massive because of the explicit chiral symmetry breaking
by quark masses.) In this paper we discuss experimental
signatures of this critical end-point.

The position of the points P and E in two-flavor QCD was estimated recently
using two different models
(a Nambu-Jona-Lasinio model
respecting the global symmetries of QCD \cite{bergesraj}
and a random matrix model \cite{stephetal}) as $T_P\sim 100$ MeV
and $\mu_P\sim 600-700$ MeV.
These are only crude estimates, since they are based
on modeling the dynamics of chiral symmetry breaking only.

The third (strange) quark has an important effect on the position
of the point P and, therefore, of the point E. 
At $\mu=0$, if the strange quark mass $m_s$ is less
than some critical value $m_{s3}$,
the second order finite $T$ transition becomes first
order.  This leads to a tricritical point in the $Tm_s$ plane
\cite{rajwil,ggp,rajreview}.  Theoretically, the origin of this
point is similar to the one we are discussing. In terms
of eq. (\ref{phi6}) the effect of decreasing $m_s$
is similar to the effect of increasing $\mu$: the coefficient $b$
becomes negative.  
What is important is that,
unlike $m_s$,
$\mu$ is a parameter which can be {\em experimentally} varied.

Clearly, the physics of the $T\mu$ plane is as in Fig. \ref{fig:pd} only
for $m_s>m_{s3}$. For $m_s<m_{s3}$, the transition
is first order already at $\mu=0$, and, presumably, remains first order
at all nonzero $\mu$ \cite{hsu}. 
As $m_s$ is reduced from infinity, the tricritical point P of Fig. \ref{fig:pd}
moves to lower $\mu$ until, at $m_s =m_{s3}$, it reaches
the $T$-axis and can be identified with the tricritical
point in the $Tm_s$ plane.   The two tricritical points are 
continuously connected.
We assume that $m_s>m_{s3}$ which is
consistent with the lattice studies of ref. \cite{columbia}. What is important
for us is that the qualitative effect of the strange
quark is to reduce the value of $\mu_P$, and thus of $\mu_E$,
compared to that in two-flavor QCD, since $\mu_P =0 $ at
$m_s=m_{s3}$.  This
shift may be significant, since lattice studies show that
the physical value of $m_s$ is of the order of $m_{s3}$.

Analysis of particle abundance ratios 
in central heavy ion collisions
\cite{stachel} indicates that chemical freeze-out happens near the
phase boundary, at a chemical potential $\mu\sim 500-600$ MeV at
the AGS (11 GeV$\cdot A$), while at the SPS ($160-200$ GeV$\cdot A$) it
occurs at a significantly lower $\mu\lesssim 200$ MeV.  In view of the effect
of the strange quark just discussed, the estimated position of P and E
\cite{bergesraj,stephetal} should be shifted from
$\mu_E\sim 600-700$ MeV to lower $\mu$. Thus, it may well be
between the SPS and the AGS values of $\mu$, 
and therefore the point E may be accessible at lower energy or
non-central collisions at the SPS.

The strategy for finding the point E 
which we propose is based on the fact that this 
is a genuine critical point. Such a point is characterized
by enhanced long wavelength fluctuations which lead to
singularities in all thermodynamic observables.
In the liquid-gas phase transition in water, 
critical opalescence signals the universal
physics unique to the vicinity of the critical point.
The signatures we propose can play an analogous role in QCD.

It is important to have control parameters which can be adjusted
to vary the $\mu$ at which the system crosses the transition region,
as shown in Fig. 2. 
For example, increasing
the energy of the collision decreases this $\mu$. 
A somewhat similar 
effect may be  
achieved by increasing centrality.
A third possibility would be to slice 
each event in rapidity,
since $\mu$ will be greater at higher rapidity.
This strategy could be useful at RHIC, if E were to
lie at lower $\mu$ than is accessible at the SPS.
We will 
call the
control parameter which is varied ``$x$'', and take increasing
$x$ to mean increasing collision energy or centrality or 
decreasing rapidity.
Scanning in centrality will 
almost certainly be the easiest, since in any given run
events with all impact parameters are present.  
However, scanning in energy yields a large variation in the $\mu$ 
at which the transition is crossed, whereas 
scanning in centrality may only
provide fine tuning.

\begin{figure}
\centerline{\psfig{file=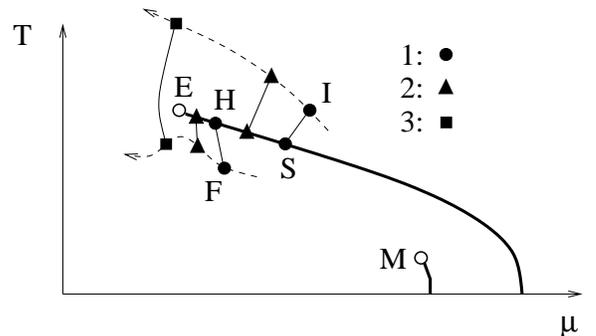,width=3in}}
\smallskip
\caption[]{Schematic examples of three possible trajectories for
three values of $x$ on the phase diagram of QCD
(see. Fig. \ref{fig:pd}). 
The points I, S, H and F on different trajectories
are marked with different symbols. The dashed lines show
the locations of the initial, I, and final, F, points
as $x$ is increased in the direction shown by the arrows.}
\label{fig:traj}
\end{figure}

In this work we do not discuss initial equilibration and we choose to
define the initial point, I$(x)$, as the point at which compression
has ended, most of the entropy is already produced, and approximately
adiabatic expansion begins.  The system will then 
follow some trajectory in the $T\mu$ plane
characterized by the ratio of the baryon charge density to the entropy 
density, $n/s$, which is (approximately) conserved.  Three trajectories
are shown schematically in Fig. \ref{fig:traj}. (For realistic 
hydrodynamical calculations
and discussion see, e.g., 
refs. \cite{Frankfurt_hydro,Hung_Shuryak}).

Recall that the first order line in the $T \mu$ plane is actually a whole
region of mixed phase, with the 
hidden parameter being the
volume fraction of the two coexisting phases.
The  zig-zag shape occurs
because the trajectory exits the mixed phase region
at a point with the same value of the conserved $n/s$ that it
had upon entering. Because $n/s$ is discontinuous at the first order 
line, this requires increasing $T$ and
decreasing $\mu$ as latent heat
is released\cite{Frankfurt_hydro,Hung_Shuryak}. 
In Fig. \ref{fig:traj}, we use the following notation: S$(x)$ 
for the ``softest'' point, H$(x)$ for
the ``hottest'' point and  F$(x)$ for the final thermal  
freeze-out after which no scattering occurs. 
(Note that at small values of $x$, 
at which the transition is first order, the trajectories
are, in fact, likely to begin within the mixed phase region.
The special case when I$(x)$ coincides with S$(x)$ leads to
a local maximum
of the QGP lifetime \cite{HS_prl,Rischke}, which may be important for
$J/\psi$ suppression \cite{ST}.)  
Increasing $x$ will yield trajectories shifted to the left in 
Fig. \ref{fig:traj},
traversing the transition region at lower $\mu$ and
higher $T$.

The existence of the end-point singularity, E, leads to the 
phenomenon which we refer to as the ``focusing'' of 
trajectories towards E.   
The initial point I$(x)$ and the beginning of the zig-zag S$(x)$
depend on the control parameter $x$ more strongly 
than the zig-zag end-point H$(x)$.
The reason for this is that the point H$(x)$ is always closer to E than
S$(x)$ (see Fig. \ref{fig:traj}).  
This focusing effect implies
that exploring physics in the vicinity of the end-point singularity
may not require a fine-tuned $x$.
This situation resembles that in low energy nuclear collisions,
in which the first order liquid-gas phase transition also has a critical
end-point at a temperature
of order 10 MeV \cite{multifragmentation1,multifragmentation2} (point
M on Figs. \ref{fig:pd} and \ref{fig:traj}).   
In such experiments,  one varies control 
parameters
to maximize the probability of multi-fragmentation.
It was noticed long ago \cite{multifragmentation1} how
surprisingly easy it is to hit the critical region
from a wide range of initial parameters.

Another aspect of the ``focusing''
arises via the divergence of susceptibilities, such as the
specific heat capacity $c_V=T\partial s/\partial T$, at the endpoint
E. As a result, the trajectories which pass near the critical point will
linger there longer.  
This makes it likely
that final freeze-out occurs at a temperature quite close to $T_E$,
rather than below it. So, while scanning in some control parameter $x$
and measuring the positions of the points F$(x)$, we may expect to find a
bump in the vicinity of the point E. (See the lower dashed curve on
Fig. \ref{fig:traj}.) 
At this point it is instructive to consider the dependence on another
control parameter, the atomic weight $A$ of the colliding nuclei.
If $A$ were infinite, the point F would be close to 
$T_F=0,\mu_F=m_N$. 
Thus, for $A$ large enough, the dotted curve 
in Fig. \ref{fig:traj} moves down and the bump, and also all
the other signatures described below, fade away.
Experimentally, the $A$ dependence of the point $F$ has been 
established recently by the analysis of flow \cite{Hung_Shuryak},
Coulomb effects\cite{Heiselberg} and pion interferometry\cite{Heinz}.
For example, in central S+S collisions at SPS $T_F\approx 140-150$
MeV, while for Pb+Pb it is only $T_F\approx 120$MeV.

We shall now discuss the signatures which
directly reflect thermodynamic properties of the system
near its critical point and are not very sensitive
to the details of the evolution.
With the advent of wide-solid-angle detectors like NA49 at CERN,
it is now possible  to make {\em event-by-event}
measurements of observables which are proxies for the
freeze-out $T$ and $\mu$  \cite{NA49}. 
We argue that the event-by-event fluctuations in 
both quantities should be anomalously small for 
values of $x$ such that the system passes near the critical
point.  As has been suggested \cite{stodolsky_shuryak},
event-by-event fluctuations of $T$ 
can be related by basic thermodynamics to
the  heat capacity at freeze-out
\begin{equation}
\frac{(\Delta T)^2}{T^2} = {1\over C_V} \ .
\label{fluctuations}
\end{equation}
The quantity $C_V$ is extensive, so $\Delta T \sim 1/\sqrt{N}$ as expected,
where $N$ is the number of particles in the system.
If the specific heat $c_V$ diverges,  the coefficient
of $1/\sqrt{N}$ vanishes and fluctuations of $T$ are suppressed.
For freezeout in the crossover region, or in the hadronic
phase just below the first order transition, $c_V$ is 
finite. (If freeze-out were to occur from the mixed phase,
some linear combination of the two susceptibilities would be relevant.)
As the critical point is approached from either the left or the right,
$c_V$ diverges and $\Delta T \sqrt{N}$ decreases.
Other susceptibilities, in particular, 
$-\partial^2 \Omega/\partial \mu^2$, are also divergent.
This implies that fluctuations of $\mu$ are also suppressed at the
critical point. Experimentally, $\Delta T$
can be found via event-by-event analysis of $p_T$ spectra
\cite{stodolsky_shuryak}.
Fluctuations in
$\mu$ correspond to event-by-event fluctuations in the 
baryon-number-to-pion ratio.
The fluctuations in any experimental observable will receive
contributions in addition to the thermodynamic ones
we describe, for example from fluctuations in the flow velocity.
We therefore expect that as the collision
energy is increased so that the freeze-out point 
moves from right to left past the critical point, 
we will find minima (but not zeroes) in the widths
of the distributions of those event-by-event
observables which are well-correlated
with $T$ and $\mu$.

Using universality, we can predict the exponents for the divergent
susceptibilities at the point E. Very naively, one might think that
the exponent describing the divergence of $C_V$ is $\alpha$, which is
small for the 3-dimensional Ising model universality class:
$\alpha\approx 0.12$.  In fact the exponent for $C_V$ is significantly
larger. This and the exponent for the $\mu$-susceptibility are
determined by finding two directions, temperature-like and
magnetic-field-like, in the $T\mu$ plane near point E,
following the standard procedure for mapping a liquid-gas
transition onto the Ising model \cite{tsypin}. 
The two linear combinations of $T-T_E$ and $\mu-\mu_E$
corresponding to these directions should then be identified (in the
sense of the universality) with the temperature, or $t=T-T_c$, and the
magnetic field, $h$, in the Ising model. One can easily understand
this by realizing that the $t$-like direction should be tangential to
the first-order line at the point E.  Then $C_V$ and $-\partial^2
\Omega/\partial \mu^2$ are different linear combinations of the
$t$-like and $h$-like susceptibilities. In both linear combinations,
the divergence of the $h$-like susceptibility will dominate because
$\gamma\approx1.2 \gg \alpha\approx0.12$.  The exponent for the
divergence of the $h$-like susceptibility as a function of the
distance, $\ell$, from the point E will depend on the direction along
which one approaches this point. For almost all directions it will be
given by $\gamma/\beta\delta\approx0.8$ (except for exactly the $t$-like
direction, where it is $\gamma$).  As a result, for points on the
$T\mu$ plane along a generic line through E one finds
\begin{equation}
(\Delta T)^2 \sim (\Delta \mu)^2 \sim \ell\,^{0.8} 
\end{equation}
sufficiently close to E.
Therefore, the fluctuations of $T$ and $\mu$ are considerably 
suppressed when the freeze-out occurs near the 
critical point.

We turn now to direct signatures of the long-wavelength
fluctuations of the massless $\sigma$ field.
For the choices of
control parameters $x$ such that freeze-out occurs at (or near)
the point E, the $\sigma$-meson is the most numerous species
at freeze-out,
because it is (nearly) massless and so 
the equilibrium occupation number of the long-wavelength modes
($T/\omega$) is large.  Because the pions are massive at the critical
point E, the $\sigma$'s cannot immediately decay into $\pi\pi$.
Instead, they persist as the density of the system further decreases. 
It is important to realize that
after freeze-out, one can (by definition) approximately neglect
collisions between particles.  Collective effects related to forward
scattering amplitudes cannot be neglected.  That is, although the
particles no longer scatter, their dispersion relations will not be
given by those in vacuum until the density is further reduced by
continued expansion.

During the expansion, the in-medium sigma mass rises towards its
vacuum value and eventually exceeds the $\pi\pi$ threshold. As the
$\sigma\pi\pi$ coupling is large, the decay proceeds
rapidly. This yields a population of pions with small transverse
momentum, $p_T < m_\pi$.  Because this process occurs after
freeze-out, the pions generated by it do not get a chance to
thermalize.  Thus, the resulting pion spectrum should have a
non-thermal enhancement at low $p_T$ which is largest
for freeze-out near E where the $\sigma$'s are most
numerous.

We now propose a specific signature of the endpoint visible
in the fluctuations of the 
pions resulting from the (formerly) long wavelength modes of the
$\sigma$ field.
For freeze-out close enough to E that the sigma mass at freeze-out
is less than $T$, the
thermal fluctuations of the number,
$N_\sigma$, of $\sigma$ particles are determined by the classical
statistics of the field $\sigma$, rather than by Poisson statistics of
particles.  Therefore, ${\langle N_\sigma^2 \rangle - \langle
N_\sigma\rangle^2 \sim \langle N_\sigma\rangle^2}$, rather than
$\langle N_\sigma\rangle$.  Thus, we expect large event-by-event
fluctuations in the multiplicity and distributions of the soft pions:
$N_\pi\approx2N_\sigma$.  Due to critical slowing down,
non-equilibrium effects may further enhance these fluctuations. Thus,
these pions could be detected either directly as an excess in the
$p_T$-spectra at low $p_T$, or via increased event-by-event
fluctuations at low $p_T$, or by an increase in Hanbury-Brown-Twiss 
(HBT) correlations
due to the larger number of pions per phase space cell at 
low~$p_T$.\cite{Shuryak_U(1)restoration}

To conclude, we propose that by varying control parameters such as
the collision energy and centrality, one may find a window of
parameters for which the $T \mu$ trajectories pass close to the
critical point E. Enhanced critical fluctuations of the $\sigma$ field
and the associated thermodynamic singularities 
lead directly 
to the signatures we propose.  When the freeze-out occurs near
the point E, we predict large non-thermal multiplicity and enhanced
event-by-event fluctuations of the soft pions.  In contrast, the
event-by-event fluctuations in both $T$ and $\mu$, as determined
using pions with $p_T \gtrsim m_\pi$, will be anomalously suppressed. 
Both effects should disappear if
the atomic weight $A$ is very large.  No one of these signatures is
distinctive in isolation and without varying control parameters.
Several of them seen together and seen to turn on and then turn off
again as a control parameter is varied monotonously would constitute a
decisive detection of the critical point.

What would we learn about QCD if such a point is found?  First, we would
learn that there is a genuine critical point in the $T\mu$ plane
in nature. Second, we would learn that $m_s>m_{s3}$ in nature,
and the $\mu=0$ thermal transition is a crossover for physical
quark masses, rather than a first-order phase transition.
Third, the 
experimental 
discovery of the critical end-point E would mean that if the light quark
masses were set to zero, there would be a tricritical point P in
the phase diagram of QCD.

We are grateful to J. Berges, M. Halasz, A. Jackson, R. Shrock and
J. Verbaarschot for fruitful collaborations 
inspiring this paper 
and helpful
conversations.  We also thank 
L. McLerran,
G. Roland and M. Tsypin for 
helpful conversations.
The work of MS is supported by NSF grant PHY97-22101.
The work of KR is supported in part by the A.P. Sloan
Foundation and by the DOE 
under agreement DE-FC02-94ER40818.
The work of ES is supported in part by DOE grant DE-FG02-88ER40388.

%\end{narrowtext}


\begin{references}

\bibitem{piswil}
R. Pisarski and F. Wilczek, Phys. Rev. {\bf D29} (1984) 338.

\bibitem{rajreview}
K. Rajagopal, Quark-Gluon Plasma 2, (World Scientific, 1995) 484, ed. R. Hwa.

\bibitem{njl}  
S.P. Klevansky, Rev. Mod. Phys. {\bf 64} (1992) 649; 
A. Barducci, R. Casalbuoni, G. Pettini and R. Gatto, Phys. Rev. {\bf D49}
(1994) 426.

\bibitem{steph} 
        M.A. Stephanov, Phys. Rev. Lett. {\bf 76} (1996) 4472;
        Nucl. Phys. B (Proc. Suppl.) {\bf 53} (1997) 469.

\bibitem{colorsuperconductor}
M. Alford, K. Rajagopal and F. Wilczek, Phys. Lett. {\bf B422} (1998) 247
and hep-ph/9804403; R. Rapp, T. Sch\"afer, E. V. Shuryak and M. Velkovsky,
hep-ph/9711396.

\bibitem{bergesraj} J. Berges and K. Rajagopal, hep-ph/9804233.

\bibitem{stephetal}
 	M.A. Halasz, A.D. Jackson, R.E. Shrock, M.A. Stephanov 
 	and J.J.M. Verbaarschot, hep-ph/9804290.

\bibitem{lawrie}
For a review, see I. Lawrie and S. Sarbach
in Phase Transitions and Critical Phenomena {\bf 9},
(Academic Press, 1984) 1, ed. C. Domb and J. Lebowitz.

\bibitem{bailin}
D. Bailin and A. Love, Phys. Rept. {\bf 107} (1984) 325.

\bibitem{rajwil}
F. Wilczek, Int. J. Mod. Phys. {\bf A7} (1992) 3911;
K. Rajagopal and F. Wilczek, Nucl. Phys. {\bf B399} (1993) 395.

\bibitem{ggp}
S. Gavin, A. Gocksch and R. Pisarski, Phys. Rev. {\bf D49} (1994) 3079.

\bibitem{hsu}
S. Hsu and M. Schwetz, hep-ph/9803386, version 3.

\bibitem{columbia}
Columbia Group, F.R. Brown et al, Phys. Rev. Lett. {\bf 65} (1990) 2491.

\bibitem{stachel}
See, e.g., P. Braun-Munziger and J. Stachel,
Nucl. Phys. {\bf A606} (1996) 320.

\bibitem{Frankfurt_hydro}
P. R. Subramanian, H. St\"ocker and W. Greiner, Phys. Lett. {\bf B173} 
(1986) 468. 

\bibitem{Hung_Shuryak}
	C. M. Hung and E. Shuryak, 
	Phys.Rev. {\bf C57} (1998) 1891.

\bibitem{HS_prl}
C. M. Hung and E. V. Shuryak, Phys. Rev. Lett. {\bf 75} (1995) 4003.

\bibitem{Rischke}
D.H. Rischke, Y. Pursun, J. A. Maruhn, H. Stocker, W.
Greiner, nucl-th/9505014. 

\bibitem{ST}
%	E. Shuryak and  D. Teaney, nucl-th/9801016, Phys. Lett. B, in press. 
	E. Shuryak and  D. Teaney, Phys. Lett. {\bf B430} (1998) 37.

\bibitem{multifragmentation1}
		L.P. Csernai and J.I. Kapusta,
		Phys. Rep. {\bf 131} (1986) 223.


\bibitem{multifragmentation2}
W. Trautmann, 
nucl-ex/9611002.

\bibitem{Heiselberg}
		H.W. Barz, J.P. Bondorf, J.J. Gaardhoje, and 
		H. Heiselberg, Phys. Rev. {\bf C57} (1998) 2536. 


\bibitem{Heinz}
U. Heinz,  Proc. of Quark Matter '97, 
nucl-th/9801050. 

\bibitem{NA49}
G. Roland for the NA49 collaboration, 
Proc. of Quark Matter '97 and Proc. of the
Hirschegg Workshop on QCD Phase Transitions, 1997.

\bibitem{stodolsky_shuryak}
L. Stodolsky, Phys. Rev. Lett. {\bf 75} (1995) 1044;
E. V. Shuryak, Phys. Lett. {\bf B423} (1998) 9.

%\bibitem{landau}
%L. D. Landau and E. M. Lifshitz, {\it Statistical Physics, Course
%of Theoretical Physics}, vol. 5, 3rd ed., (Pergamon, 1980).

%\bibitem{amit}
%D. J. Amit, {\it Field Theory, the Renormalization Group, and
%Critical Phenomena}, (World Scientific, 1984).


\bibitem{tsypin} See, e.g.,
K. Rummukainen, M. Tsypin, K. Kajantie, M. Laine, and M. Shaposhnikov,
hep-lat/9805013.


\bibitem{Shuryak_U(1)restoration} 
Note that this HBT signal is distinct from that proposed
in E. Shuryak,
Comments Nucl. Part. Phys. {\bf 21} (1994) 235 and 
J. Kapusta, D. Kharzeev and L. McLerran, Phys. Rev. {\bf D53} (1996)
5028.

%\bibitem{Kharzeev}
%S. E. Vance, T. Csorgo and D. Kharzeev, nucl-th/9802074.


\end{references}
\end{document}